\documentclass[aps,prl,twocolumn,showpacs,floatfix,superscriptaddress]{revtex4-1}

\usepackage{times}

\usepackage{amsmath}
\usepackage{amssymb}
\usepackage{physics}
\usepackage{graphicx}
\usepackage{comment}

\begin{document}


\title{Energy redistribution and spatio-temporal evolution of correlations after a sudden quench of the Bose-Hubbard model}





\author{Yosuke Takasu}
\affiliation{Department of Physics, Graduate School of Science, Kyoto University, Kyoto 606-8502, Japan}
\author{Tomoya Yagami}
\affiliation{Department of Physics, Graduate School of Science, Kyoto University, Kyoto 606-8502, Japan}
\author{Hiroto Asaka}
\affiliation{Department of Physics, Graduate School of Science, Kyoto University, Kyoto 606-8502, Japan}
\author{Yoshiaki Fukushima}
\affiliation{Department of Physics, Graduate School of Science, Kyoto University, Kyoto 606-8502, Japan}
\author{Kazuma Nagao}
\affiliation{Yukawa Institute for Theoretical Physics, Kyoto University, Kyoto 606-8502 Japan}
\affiliation{Zentrum f\"ur Optische Quantentechnologien and Institut f\"ur Laserphysik, Universit\"at Hamburg, 22761 Hamburg, Germany}
\affiliation{The Hamburg Center for Ultrafast Imaging, Luruper Chaussee 149, 22761 Hamburg, Germany}
\author{Shimpei Goto}
\affiliation{Department of Physics, Kindai University, 3-4-1 Kowakae, Higashi-Osaka, Osaka 577-8502, Japan}
\author{Ippei Danshita}
\affiliation{Department of Physics, Kindai University, 3-4-1 Kowakae, Higashi-Osaka, Osaka 577-8502, Japan}
\author{Yoshiro Takahashi}
\affiliation{Department of Physics, Graduate School of Science, Kyoto University, Kyoto 606-8502, Japan}









\begin{abstract}
An optical-lattice quantum simulator is an ideal experimental platform to investigate non-equilibrium dynamics of a quantum many-body system, which is in general hard to simulate with classical computers.
Here, we use our quantum simulator of the Bose-Hubbard model to study dynamics far from equilibrium after a quantum quench.
We successfully confirm the energy conservation law in the one- and three-dimensional systems and extract the propagation velocity of the single-particle correlation in the one- and two-dimensional systems.
We corroborate the validity of our quantum simulator through quantitative comparisons between the experiments and the exact numerical calculations in one dimension.
In the computationally hard cases of two or three dimensions, by using the quantum-simulation results as references, we examine the performance of a numerical method, namely the truncated Wigner approximation, revealing its usefulness and limitation.
This work constitutes an exemplary case for the usage of analog quantum simulators.
\end{abstract}

\maketitle


\section*{Introduction}

Rapid advances in analog quantum simulation using highly controllable systems with long coherence time, such as ultracold gases in optical lattices (for example, see ref.\cite{gross-17, hofstetter-18}), Rydberg atoms in an optical tweezer array (for example, see ref.~\cite{bernien-17, barredo-18, keesling-19}), and trapped ions (for example, see ref.~\cite{blatt-12, garttner-17}),
have significantly expanded possibilities for studying dynamics of quantum many-body systems. 
One of the recent targets of optical-lattice quantum simulators has been the investigation of the non-equilibrium  dynamics arising after a quantum quench~\cite{winkler-06, wirth-10, trotzky-12, cheneau-12, braun-15, greiner-02, chen-11, meinert-13, islam-15, clark-16}, where a parameter of the system is varied rapidly and substantially.
In the case of one dimension (1D) for a short time scale, quantum quench dynamics can be exactly computed with classical computers by means of the matrix product state (MPS) method (for example, see ref.~\cite{schollwock-2011,paeckel-19}).
In pioneering works of quantum-simulation research, the outputs of experiments were directly compared with those of exact numerical simulations with classical computers in order to examine the performance of the quantum simulators~\cite{trotzky-12, cheneau-12, braun-15}.

A two-point spatial correlation as a function of the distance of the two points has been the intense theoretical interest~\cite{lauchli-08, barmettler-12, natu-13, carleo-14, song-14, strand-15, bohrdt-17, fitzpatrick-18, cevolani-18, nagao-19, despres-19}, and in fact, in one-dimensional systems, it has been shown that access to such a correlation function allows for exploring the dynamical spreading of quantum information, which is of great interest in connection with the Lieb-Robinson (LR) bound~\cite{cheneau-12, jurcevic-14}.
An exact computation of the spatio-temporal evolution of such two-point correlations is, however, generally intractable for a long time scale or in higher dimensions. 
While a more recent work has utilized outputs from a quantum simulator built with ultracold fermions in a Floquet-engineered  optical lattice in three dimensions (3D) as a reference for examining the performance of an approximate numerical method, namely, the non-equilibrium dynamical mean-field theory~\cite{sandholzer-18}, a direct comparison with quantitative theoretical approaches in the quench dynamics in higher dimensions is still lacking.

In this paper, we investigate the energy redistribution dynamics and the spatio-temporal evolution of the single-particle correlation function, which is one of the simplest two-point spatial correlations, in quantum quench dynamics starting with a Mott insulating state by using an optical-lattice quantum simulator of the Bose-Hubbard model (BHM) in two dimensions (2D) and 3D as well as 1D.
%
%
The observation of the redistribution of the kinetic- and interaction-energies turns out to be the confirmation of the energy conservation in the quench dynamics of a Bose-Hubbard quantum simulator.
Further, we successfully observe the correlation spreading after a rapid quench from a Mott insulating state towards the quantum critical region in 2D as well as towards the Mott region in 1D. 
We compare the measured propagation velocity of the correlation front, which is defined from the first peak in the time evolution of the correlation function at each distance, with the LR-like bound set by the maximum velocity of the quasi-particles. In the 2D case, we find that the former velocity exceeds the latter one. This happens because the single-particle correlation spreads with two typical velocities, namely the group velocity and the phase velocity, as was pointed out in the recent theoretical work~\cite{despres-19}, and the measured velocity corresponds to the phase one. Since the first peak propagating with the phase velocity decays rapidly with the distance, our observation is not contradicting the existence of the LR bound implying that any correlation functions outside the LR light cone must be exponentially suppressed.

In addition to these experimental findings, in order to corroborate the quantitative performance of our quantum simulator, we present a thorough comparison between the quantum-simulation results and state-of-the-art theoretical calculations.
We employ the exact MPS method in the 1D case, finding excellent agreement with the observations.
As for the case of the quench towards a deep superfluid region in 3D, the time evolution of the kinetic and interaction energies is directly compared with numerical results obtained using the truncated Wigner approximation (TWA) based on the Gross-Pitaevskii mean-field theory~\cite{nagao-19}.
The good agreement between the experiment and the theory establishes the predictive power of the TWA for this type of quench.
In contrast, in the case of the quench towards the quantum critical region in 2D, the TWA fails to capture quantitatively the experimental results, although it captures some qualitative features.
This indicates that our quantum simulation goes beyond current classical computation and the data serves as a useful reference for pushing out its boundary.


\section*{Results}
\subsection*{Investigating non-equilibrium dynamics of the Bose-Hubbard model}

We consider a system of ultracold bosonic atoms confined in an optical lattice.
When an optical lattice potential is deep, the system is quantitatively described by the BHM \cite{fisher-89, jaksch-98},
\begin{multline}
\hat{\mathcal{H}}=-J\sum_{\langle j, l \rangle}\left(\hat{a}^{\dagger}_j \hat{a}_l +h.c.\right)+\frac{U}{2}\sum_{j} \hat{a}^{\dagger}_j\hat{a}^{\dagger}_j\hat{a}_j\hat{a}_j  \\
+\sum_{j}(V_{j}-\mu)\hat{a}^{\dagger}_j\hat{a}_j, ~\label{eq:Hamiltonian}
\end{multline}
where $\hat{a}^{\dagger}_j$ and $\hat{a}_j$ are the creation and annihilation operators at the site $j$, $J$ is the tunneling-matrix element between nearest-neighbor sites, $U$ is the on-site interaction energy, $\mu$ is the chemical potential, and $V_j$ is the local potential offset at the site $j$, which originates from the trap potential and the gaussian envelopes of optical lattice lasers.
$\sum_{\langle j, l\rangle}$ represents the summation over all neighboring sites.
The position of the site $j$ is denoted by ${\boldsymbol r}_j = \sum_{\alpha=1}^{D}x_j^{\alpha} {\boldsymbol e}_{\alpha}$, where $x^{1}$, $x^{2}$, and $x^{3}$ mean $x$, $y$, and $z$, respectively. ${\boldsymbol e}_{\alpha}$ represents the unit vector in the $x^{\alpha}$-direction and $D$ is the spatial dimension.

When the atom number per site, namely the filling factor $\bar{n}$, is an integer and the ratio $U/J$ is varied, the BHM exhibits a second-order quantum phase transition between the Mott insulator and the superfluid.
The system favors the superfluid phase for a relatively small $U/J$ while it does the Mott insulator phase for a relatively large $U/J$. For the unit filling case ($\bar{n}=1$), the quantum critical point has been determined with exact numerical methods as $\left(U/J\right)_c =$ 3.4 (1D), 16.7 (2D), and 29.3 (3D), respectively (for review, see ref.~\cite{krutitsky-16}).

Our analog quantum simulator of the BHM is built with an ultracold Bose gas of ${}^{174}\text{Yb}$ atoms confined in a 3D optical lattice.
We use this ${}^{174}\text{Yb}$-atom-BHM quantum simulator in order to analyze dynamics after a quench of the ratio $U/J$ starting with a Mott-insulator state with unit filling.
We convert a ${}^{174}\text{Yb}$ Bose-Einstein condensate (BEC) in a weakly confining harmonic trap into the initial Mott-insulator state by slowly ramping up the optical-lattice depth up to $s\equiv V_0/E_{\rm R} = 15$ for all the three directions, where $V_0$ is the depth of the optical lattice and $E_{\rm R}$ is the recoil energy of the optical-lattice laser whose wavelength is 532 nm.  
See Methods for the preparation.
The prepared state is deep in a Mott insulator regime ($U/J=100$) and is well approximated as a product of local Fock states, 
\begin{eqnarray}
\ket{\Psi_{\rm MI}}=\prod_j\hat{a}_j^{\dagger} \ket{0}. 
\label{eq:deepMott}
\end{eqnarray}


To realize a quench of $U/J$, we rapidly ramp down the lattice depth for some directions towards a final value. For instance, in the case of the 1D quench we ramp down the lattice depth only for the $x$ direction while in the 3D case we do it for all the three directions. The ramp-down speed is set to be 100 $E_{\rm R}/{\rm ms}$. 
By using the band-mapping techniques, we check that there is no discernible amount of the atoms in excited bands with this quench speed.
We use the numerical values of $U$ and $J$ calculated as functions of lattice depth reported in ref.~\cite{krutitsky-16} (See Section I of Supplementary Materials).

After the quench process, we keep the lattice depth constant and let the system evolve.
In order to obtain the single-particle correlation function at a certain distance ${\boldsymbol \Delta} = \sum_{\alpha=1}^D
{\boldsymbol e}_{\alpha} \Delta_{x^{\alpha}}$, where $\Delta_{x^{\alpha}}\geq 0$, in the unit of a lattice spacing $d$ (=266 nm)\cite{nakamura-19}, 
\begin{equation}
K_{\boldsymbol \Delta} = 
\sum_{\alpha = 1}^{D}\sum_{|x_j^{\alpha}-x_l^{\alpha}|=d\Delta_{x^\alpha}}
\langle
\hat{a}_j^{\dagger}\hat{a}_l
\rangle \label{eq:K}
\end{equation}
after a certain hold time, we release the gas from the trapping and optical-lattice potentials to measure the time-of-flight (TOF) image, from which we deduce the momentum distribution (See Methods for details).
By performing a Fourier transform of the momentum distribution, we obtain $K_{\boldsymbol \Delta}$.
The kinetic energy of the BHM is equal to the sum of $-JK_{\boldsymbol \Delta}$ at ${\Delta}=1$, where $\Delta = |{\boldsymbol \Delta}|$\cite{nakamura-19}.
Moreover, we measure the onsite-interaction energy of the BHM, $\frac{U}{2}\sum_{j} \langle\hat{a}_j^{\dagger}\hat{a}_j^{\dagger}\hat{a}_j\hat{a}_j\rangle$, by means of the atom-number-projection spectroscopy \cite{kato-16, nakamura-19} for the 3D case and photoassociation spectroscopy~\cite{sugawa-11} for the 1D case where we confirm that there are almost no multiple occupancies larger than two (See Methods for details).
Compared to the methods based on the quantum-gas microscope techniques~\cite{cheneau-12}, our methods are rather efficient, especially in higher dimensions, for our current purposes of obtaining the ensemble average of the two-point correlation functions and the Hubbard energies, because less repetitions are needed thanks to the much larger number of atoms.
The experimental procedure and set-up, and typical high-resolution spectra are summarized in Sections I and II of the Supplementary Materials. 
It is worth noting that the dynamical evolution of the phase correlation, which is similar to the single-particle correlation, has been measured for weakly interacting Bose gases in one-dimensional optical lattices by means of Talbot interferometry in ref.~\cite{santra-17}. In contrast, the present work investigates the single-particle correlation in strongly correlated regimes in higher dimensions.

\subsection*{Experimental confirmation of our methods: Dynamics of the 1D Bose Hubbard model after a sudden quench}

\begin{figure}
	\centering
	\includegraphics[width=7cm]{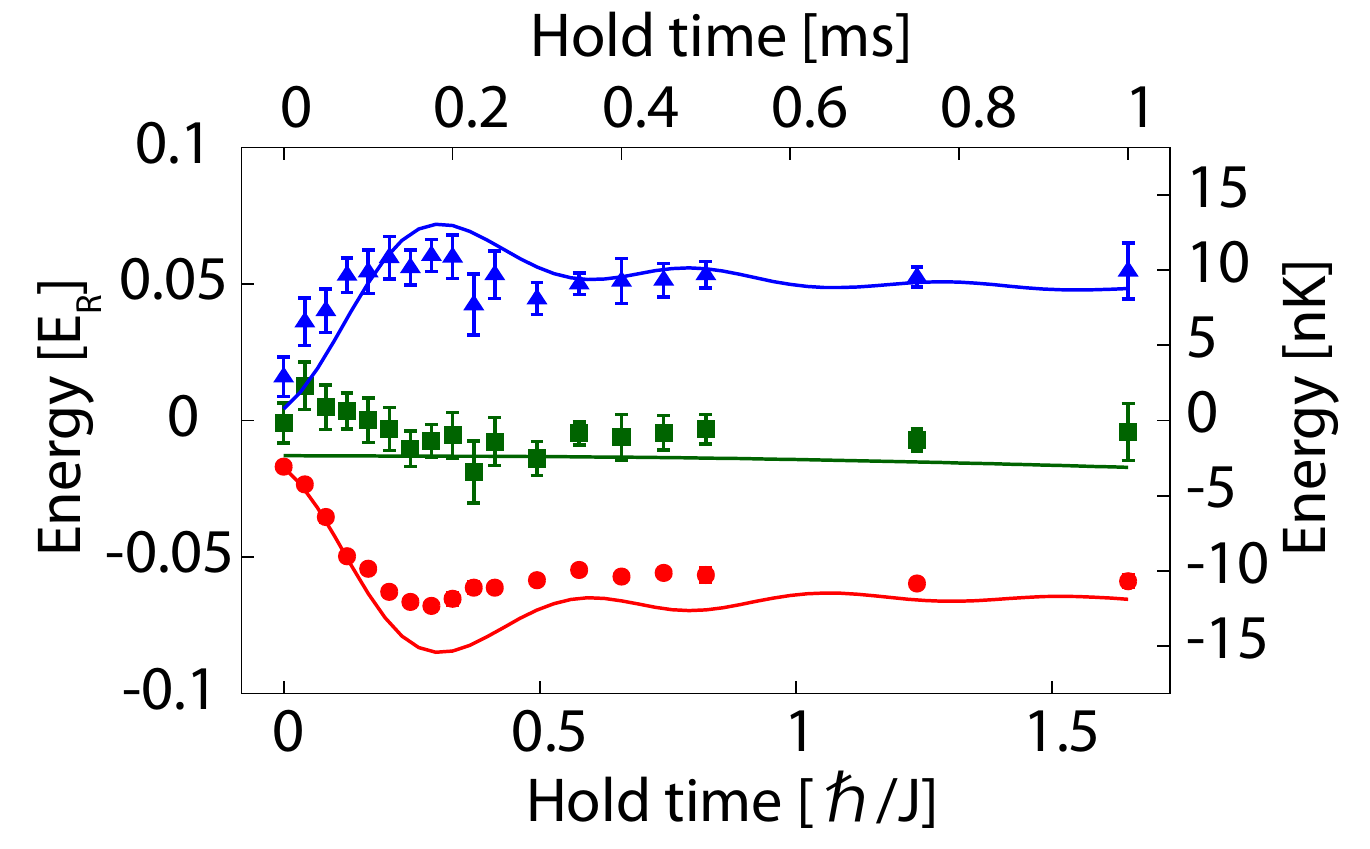}
	\caption{{\bf Energy redistribution after the quench in one dimension.} The kinetic-energy term (red), the onsite-interaction-energy term (blue), and the sum of them (green) are shown as functions of the hold time $t$ after a rapid quench into a Mott insulator region with $U/J=6.8$ in 1D optical lattice tubes. The solid lines show the results of the numerical calculation at zero temperature with the MPS method using the time-dependent variational principle and local density approximation(LDA).\label{fig:1Dquench}
		The error bars for the kinetic-energy and onsite-interaction-energy terms denote the standard error of 15 independent measurements.}
\end{figure}

\begin{figure*}
	\centering
	\includegraphics[width=14cm]{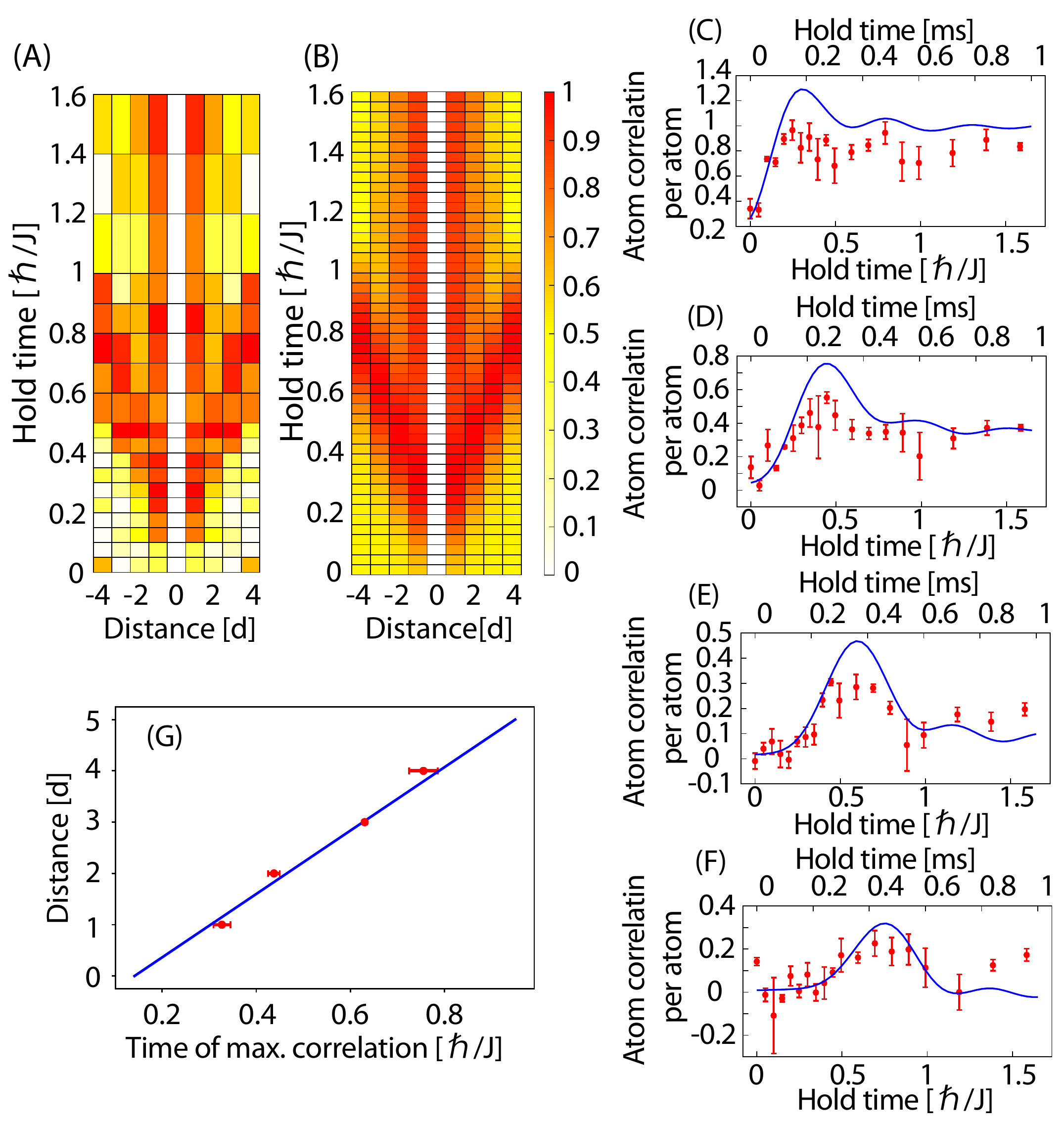}
	\caption{{\bf Spatio-temporal evolution of the single-particle correlation after the quench in one dimension.} 
		({\bf A}), ({\bf B}) The single-particle correlations for the distance in the unit of the lattice constant $\Delta$ up to 4 are shown as functions of the hold time $t$. Note that the displayed correlations are normalized by the maximum value of the correlation $C^{(1D)}_{{\rm max},\Delta}$ during $0<t<1.6 \hbar/J$ for each distance $\Delta$; ({\bf A}), Experiment; ({\bf B}), numerical calculation.
		({\bf C})-({\bf F}), Time evolution of the single-particle correlation $K_\Delta$ after the quench. Solid blue lines show the results of numerical calculation. ({\bf C}), $\Delta=1$; ({\bf D}), $\Delta=2$; ({\bf E}), $\Delta=3$; ({\bf F}), $\Delta=4$. 
		The error bars denote the standard error of 5 independent measurements. 
		({\bf G}) Time of the first peak of the single-particle correlation is plotted as a function of the distance $\Delta$.
		A fit with a linear function with a non-zero offset is shown as a solid line.
		The error bars denote the standard error of 5 independent measurements. 
		\label{fig:1Dcorrelation}}
\end{figure*}

First, we investigate the behaviors of atoms after a sudden quench in 1D.
The results for the dynamical redistribution of the Hubbard energies and the spatio-temporal evolution of the atom correlations are shown in Figs.~\ref{fig:1Dquench} and \ref{fig:1Dcorrelation}, respectively.
Specifically, we ramp the lattice depth in the $x$ direction down to $s = 5$ implying $U/J=6.8$, where the ground state is a Mott insulator state close to the quantum critical point.
Figure~\ref{fig:1Dquench} shows the time evolution of the kinetic energy, the onsite-interaction energy, and the sum of the two energies. 
On a short time scale, while the sum of the two remains almost constant, the kinetic energy decreases and the interaction energy increases. 
After making a small overshoot, each energy ends up with an almost steady value, i.e., the energies are redistributed. 
These behaviors are expected for an isolated system but have never been observed experimentally before.

Figures ~\ref{fig:1Dcorrelation}A-\ref{fig:1Dcorrelation}D show the time evolution of the single-particle correlation function for several values of $\Delta$. 
As the time evolves, the correlations first grow and each of them has the first (local) maximum at a certain time. We extract the peak time for each $\Delta$ by numerically fitting to the experimental data, which is plotted against $\Delta$ in Fig.~\ref{fig:1Dcorrelation}G. The peak time increases linearly with the distance, i.e., the correlation exhibits a light-cone-like propagation. From the peak time versus $\Delta$, we extract the propagation velocity as $v = 5.5(7) Jd/\hbar$.
The maximum velocity of a particle-hole excitation is given as\cite{cheneau-12, barmettler-12}
\begin{equation}
v_{\text{max}} \simeq \frac{6J d\sqrt{D}}{\hbar}\left[1-\frac{16J^2}{9U^2}\right], \label{eq:LR}
\end{equation}
which can be interpreted as a LR-like bound.
It is noted that $v_{\rm max}$ corresponds to the sum of the maximum velocities of the doublon and the holon, which are respectively given by $4Jd\sqrt{D}/\hbar$ and $2Jd\sqrt{D}/\hbar$ in the leading order with respect to $J/U$. As long as $U/J \gg 1$, Eq.~(\ref{eq:LR}) is valid regardless of the spatial dimension.
At $U/J = 6.8$, $v_{\rm max} = 5.8 Jd/\hbar$ such that the condition $v < v_{\rm max}$ is satisfied, as expected.
A similar propagation behavior has been also observed in the case of the density-density correlation \cite{cheneau-12}.
In contrast, we will see later that $v>v_{\rm max}$ in the 2D case. We will explain that this observation is still compatible with the LR-like bound.

While these observations reveal important features of the non-equilibrium dynamics of BHM, this 1D study is also important from another aspect.
Since our BHM quantum simulator is analog, it is imperative to examine its accuracy through a direct comparison with exact numerical calculations in 1D before applying it to the cases of higher dimensions, in which exact computation on classical computers is currently unavailable.
In Figs.~\ref{fig:1Dquench} and \ref{fig:1Dcorrelation}, we compare the experimental results in 1D with the exact numerical ones at zero temperature obtained with the MPS method.
For details of the MPS calculations, see Sec. IV of the Supplementary Materials.
We see that the experimental observations are in good agreement with the exact numerical calculations with no fitting parameters.

\subsection*{Dynamics of BHM after a sudden quench in higher dimensions}

\begin{figure}
	\centering
	\includegraphics[width=7cm]{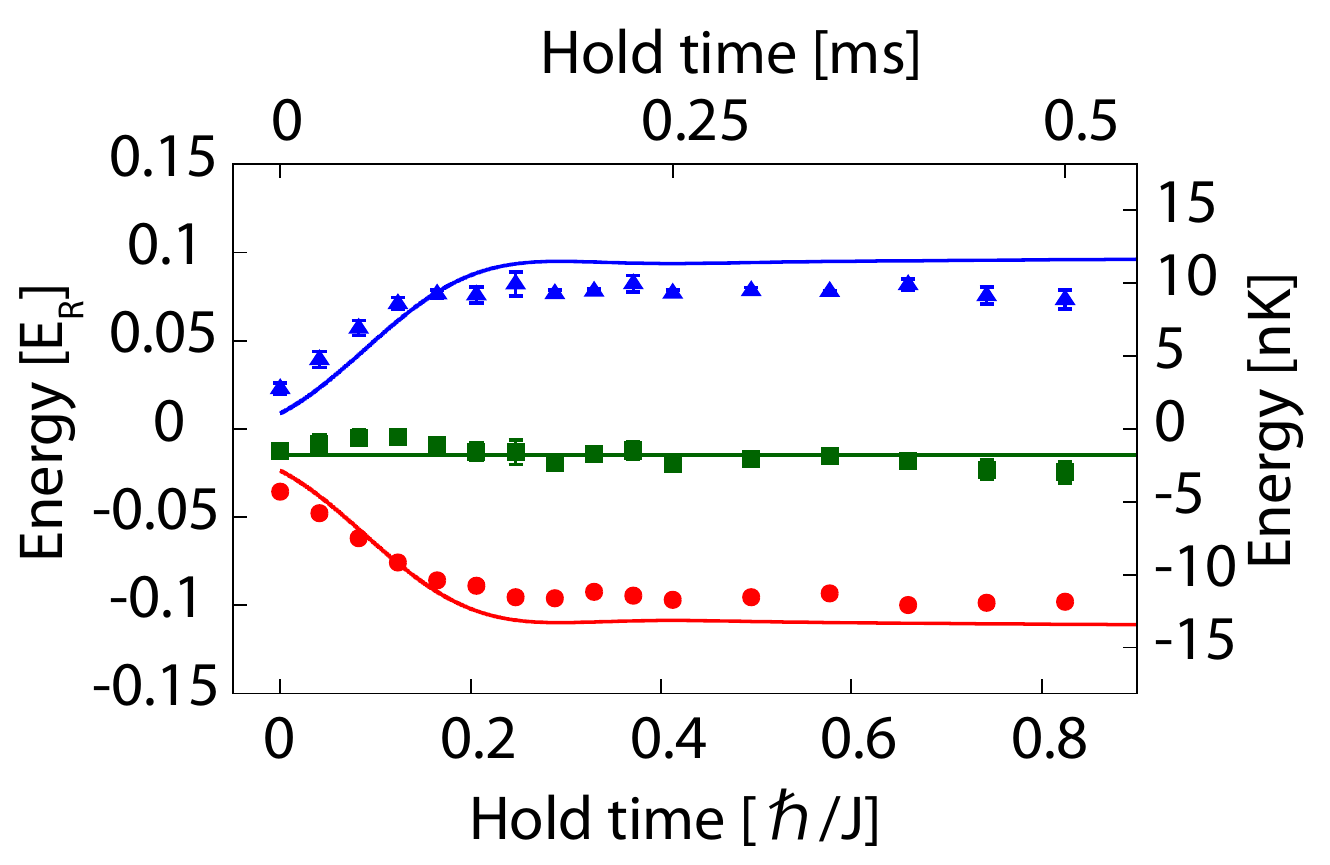}
	\caption{{\bf Energy redistribution after the quench in three dimensions.} 
		The kinetic-energy term (red), the onsite-interaction-energy term (blue), and the sum of them (green) are shown as functions of the hold time $t$ after a rapid quench into a superfluid region with $U/J = 3.4$ in a 3D optical lattice. 
		The solid lines show the results of the numerical calculation obtained using the TWA~\cite{nagao-19}\label{fig:3Dquench}.
		The error bars for the kinetic-energy (onsite-interaction energy) terms denote the standard error of 15 (3) independent measurements.
	}
\end{figure}

Having corroborated the quantitative validity of our BHM quantum simulator by the comparison between the theory and experiment in 1D, we now discuss the main result of this work, i.e., the quench dynamics in higher dimensions. 
Figure~\ref{fig:3Dquench} shows the energy-redistribution dynamics for the 3D case after the ramp-down of the lattice depth to $s = 5$ ($U/J = 3.4$), where the ground state is deep in the superfluid phase. The general tendency of the time evolution is similar to the 1D case: the two energies are redistributed on a time scale smaller than $\hbar/J$ and the sum of the two remains almost constant within the displayed time window $t\lesssim\hbar/J$.

\begin{figure*}
	\centering	
	\includegraphics[width=14cm]{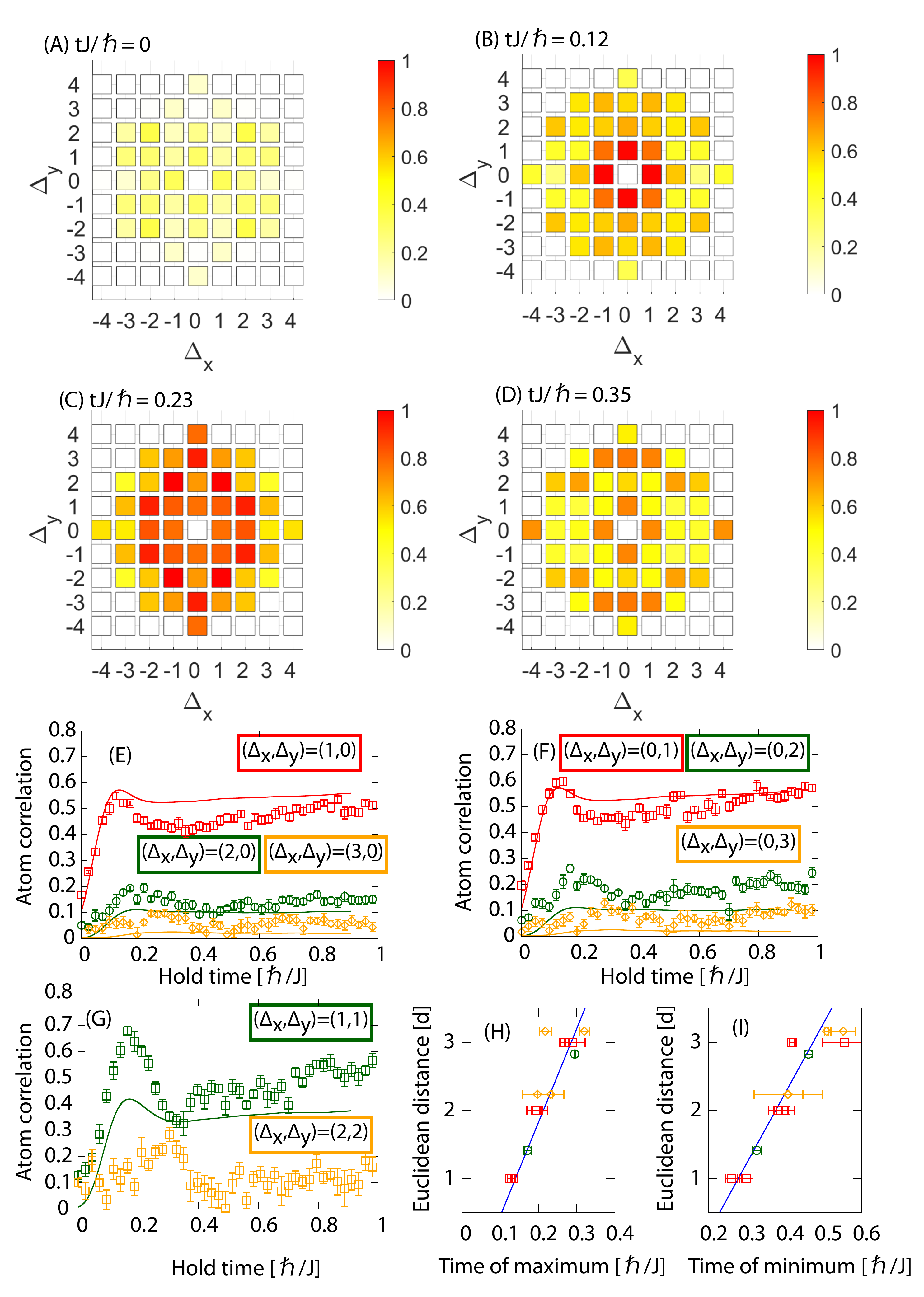}
	\caption{{\bf Spatio-temporal evolution of the single-particle correlation after the quench in two dimensions.} 
		({\bf A})-({\bf D}) 2D plots of the single-particle correlation as functions of the distances $\Delta_x$ and $\Delta_y$ for several hold times of $tJ/\hbar$= 0 ({\bf A}), 0.12 ({\bf B}), 0.23 ({\bf D}), and 0.35 ({\bf D}). 
		Data with $\sqrt{\Delta_x^2+\Delta_y^2}\le4$ are shown.
		Note that the displayed correlations are normalized by the maximum value of the correlation $C^{(2D)}_{{\rm max},{\boldsymbol \Delta}}$ during $0<t<1.0 \hbar/J$ for each distance ($\Delta_x$, $\Delta_y$).
		({\bf E})-({\bf G}) Time evolution of the single-particle correlation $K_{\boldsymbol \Delta}$ after the quench for ($\Delta_x$, $\Delta_y$) = (1,0) (({\bf E}), red square), (2,0) (({\bf E}), green circle), (3,0) (({\bf E}), yellow diamond), (0,1) (({\bf F}), red square), (0,2) (({\bf F}), green circle), (0,3) (({\bf F}), yellow diamond), (1,1) (({\bf G}), green circle), and (2,2) (({\bf G}), yellow diamond).
		The solid lines are the numerical results obtained using the TWA method. 
		The error bars denote the standard error of 15 independent measurements.
		({\bf H})-({\bf I}) Time at the first peak ({\bf H}) or the first trough ({\bf I}) of the single-particle correlation as a function of the Euclidean distance $\Delta=\sqrt{\Delta_x^2+\Delta_y^2}$. 
		A fit with a linear function with a non-zero offset is shown as a solid line both in ({\bf H}) and ({\bf I}).
		The first peak and trough are obtained by fitting the experimental data with the empirical function described in Sec VI of Supplementary Materials. 
		The error bars denote the fitting errors.
		\label{fig:2Dcorrelation}}
\end{figure*}

We next investigate the dynamical spreading of the single-particle correlation after a quantum quench in 2D.
The final lattice depth in this case is $s = 9$ implying $U/J=19.6$, where the ground state is a Mott insulator phase near the quantum critical point.
Figures~\ref{fig:2Dcorrelation}A-~\ref{fig:2Dcorrelation}D show the spatial distribution of the single-particle correlation at several hold times after a quench. We clearly observe that the correlation first grows between nearest-neighbor sites and then it propagates for larger distances at later times.
More directly, Figs.~\ref{fig:2Dcorrelation}E-\ref{fig:2Dcorrelation}G show the time evolution of the single-particle correlation function $K_{\Delta_x,\Delta_y}$ for several values of $(\Delta_x,\Delta_y)$, where $\Delta_x$ ($\Delta_y$) denotes the distance in the $x$ ($y$) direction in units of the lattice spacing $d$.
The delay in the growth of the correlation for longer distance is clearly observed along the directions of $x$, $y$, and $x+y$, in Figs.~\ref{fig:2Dcorrelation}E, \ref{fig:2Dcorrelation}F, and \ref{fig:2Dcorrelation}G, respectively.
In the same manner as the 1D case, we extract the position of the first peak in the time evolution of the correlation at each distance, which is plotted against the Euclidean distance $\sqrt{\Delta_x^2 + \Delta_y^2}$ in Fig.~\ref{fig:2Dcorrelation}H.
We further extract the propagation velocities from the linear fitting to Figs.~\ref{fig:2Dcorrelation}H and \ref{fig:2Dcorrelation}I as $v=13.7(2.1) Jd/\hbar$ (peak) and $v=10.2(1.4)Jd/\hbar$ (trough). 
According to Eq.~(\ref{eq:K}), the maximum velocity of the particle-hole excitation is $v_{\rm max} = 8.4 Jd/\hbar$, which is slower than the observed propagation velocity in Fig.~\ref{fig:2Dcorrelation}.

\section*{Discussion and Outlook}

The observation that $v>v_{\rm max}$ in the 2D case shown in Fig.~\ref{fig:2Dcorrelation} can be interpreted along the line explained in ref.~\cite{despres-19}. The correlation spatially propagates as a wave packet, whose width spreads in time.
This means that the velocity of the first peak in the time evolution of the correlation function at each distance, namely the phase velocity, is faster than that of the center of the wave packet, namely the group velocity. Moreover, the first peak decays rather rapidly as the distance becomes larger. The velocity extracted from the experimental data in the way described above corresponds to the phase velocity while the meaningful propagation velocity, which should be compared with the LR-like bound, does to the group velocity. In the case of the final lattice depth $s=9$ in 2D, we cannot accurately extract the group velocity because of the unclear separation of the two velocities. Instead, In Sec. V of the Supplementary Materials, we show an example, in which the phase velocity is well separated from the group velocity in the 1D case with large $U/J$. There we also see that $v>v_{\rm max}$. Hence, the behavior that $v>v_{\rm max}$ is not unique to the 2D case but can emerge regardless of the spatial dimension as long as $U/J$ after the quench is sufficiently large. Notice that we observed $v<v_{\rm max}$ in the case of the final lattice depth $s=5$ ($U/J = 6.8$) in 1D because the phase velocity is approximately equal to the group velocity at $U/J = 6.8$~\cite{despres-19}.

Next, we discuss the usefulness and limitation of some numerical methods based on the quantum simulation results. Since there is no exact computation method applicable to the 2D and 3D cases, it is meaningful to examine the accuracy of some approximate methods by using the quantum simulation results as a quantitative reference. In ref.~\cite{sandholzer-18} the time-dependent dynamical mean-field theory has been examined by comparison with quantum simulation results for real-time dynamics of the Fermi-Hubbard model. This method is not suited for computing the non-local spatial correlations analyzed in the present work because it ignores the momentum dependence of the correlation functions. Instead, we choose the TWA approximation, which is supposed to accurately capture semiclassical dynamics of the BHM at least on a short time scale (See ref.~\cite{nagao-19} and references therein).
In Fig.~\ref{fig:3Dquench}, where the energy-redistribution dynamics in 3D is depicted, we also show the numerical calculations as solid lines obtained with the TWA~\cite{nagao-19}. 
In the TWA calculations, we take the Mott insulator state of Eq. (\ref{eq:deepMott}) as the initial state and set the system size to be $30^3$ sites. We ignore the trapping potential because it is irrelevant to the dynamics within the time window $t\lesssim\hbar/J$ as was discussed in the 1D case.
The TWA results are in good agreement with the experimental observations.
More details of the TWA calculations are described in ref.~\cite{nagao-19}.

Let us turn our attention to the correlation spreading in 2D shown in Fig.~\ref{fig:2Dcorrelation}.
The solid lines in Figs.~\ref{fig:2Dcorrelation}E-\ref{fig:2Dcorrelation}G represent the results obtained by using the TWA.
The TWA agrees with the experiment on a very short time scale ($t<0.1\hbar/J$). Moreover, the peak positions and the values at a relatively long time ($t > 1\hbar/J$) for a short distance, say $\Delta = 1$, are reasonably captured. However, it fails to capture some important properties of the correlation dynamics, such as the locations of the correlation troughs and the almost converged value of the correlation for $\Delta > 1$. This disagreement is consistent with the general fact that the TWA is less accurate when $U/(D\bar{n}J)$ or $tJ/\hbar$ is larger. This failure of the TWA indicates that one needs to push out the boundary of currently available numerical techniques for quantitative description of the quantum simulation results. One possible candidate is to extend the SU($N$) TWA~\cite{schachenmayer-15, davidson-15} for analyzing the BHM with unit filling. 

In both of the 1D and 2D quench cases, we observed that the peaks propagated linearly with a constant velocity (see Fig.~\ref{fig:1Dcorrelation}G and Figs.~\ref{fig:2Dcorrelation}H,\ref{fig:2Dcorrelation}I).
However, extrapolations to $t=0$ have non-zero offsets.
In addition, our numerical results also support the existence of the offsets.
The offsets reflect the difference between the speed for the creation of a particle-hole pair and that for its propagation. The former speed determines the time giving the first peak at $\Delta=1$ while the latter does those at $\Delta >1$.
It is noted that the dependence of the propagation velocity on distance in the case of the 1D quench was already numerically discussed in ref.~\cite{cheneau-12}.
Our quantum simulation platform for studying non-equilibrium dynamics can be straightforwardly applied to other quantum many-body systems such as the Fermi-Hubbard model (with SU(N) symmetry~\cite{zhang-19, huang-19}), the Bose-Fermi Hubbard model, and the spinful BHM. In addition, it is interesting to extend our work to a study of quench dynamics on a quantum system with controlled dissipation, which has recently attracted much interest~\cite{tomita-17}.
%


\section*{Methods}

\subsection*{Preparation of initial Fock state}
Details of our experimental setup are described in ref.~\cite{nakamura-19}.
We first prepare a BEC of ${}^{174}\text{Yb}$ atoms confined in an optical far-off resonant trap (FORT) whose wavelength is 532 nm.
The trap frequencies of the FORT are given by $(\omega_{x’}, \omega_{y'},\omega_z) = 2\pi \times (28,130,160)$ Hz, where the $x’$ and $y’$ axes were tilted from the $x$ and $y$ axes, to which two of the optical lattices are directed, by 45$^{\circ}$.
Then we slowly ramp up the optical-lattice depth for all the three directions from $s= 0$ to $5$ in 100 ms and from 5 to 15 in another 100 ms.
A typical number of atoms is chosen to be $N = 1.3\times 10^4$ such that the filling factor is unity.

\subsection*{Lattice quench}
We perform the quench by sudden decrease of the optical lattice with depth of $s$ $E_{\rm R}$ in $0.01(15-s)$ ms.
See also Sec. I of Supplementary Materials.
The excitation of the atoms into higher bands is negligible with this procedure.
For the cases of the 1D and 2D quench, we decrease the lattice depth along the one direction of $x$ and two directions of $x$ and $y$, respectively.
It is to be noted that when lattice depth is 10.6 $E_{\rm R}$, $U/J$ is equal to $29.34$, which is the critical  lattice depth for the superfluid-Mott transition at $\bar{n}=1$.

\subsection*{Measurement of the ensemble average of the non-local atom correlation}

Here we briefly describe a method for obtaining the ensemble average of the non-local atom correlation $K_{\boldsymbol \Delta}$ of Eq.~(\ref{eq:K}).
Details are described in ref.~\cite{nakamura-19}.
The atomic-density distribution $n (\mathbf{r})$ after the TOF $t$ is given by
\begin{equation}
n(\mathbf{r})=\left(\frac{m}{\hbar t}\right)^3 \left|\tilde{w}_0(\mathbf{k})\right|^2 S(\mathbf{k}),
\end{equation}
where $\tilde{w}_0 (\mathbf{k})$ is the Fourier transformation of the Wannier function in the lowest Bloch band $w_0(\mathbf{r})$, and $\mathbf{k}$ = $m\mathbf{r}/\hbar t$.

When $t$ is long enough and the structure factor $S(\mathbf{k})$ is expressed as
%
\begin{equation}
S(\mathbf{k})=\sum_{j,l}e^{i \mathbf{k}\cdot\left(\mathbf{r}_j-\mathbf{r}_l\right)} \langle \hat{a}^{\dagger}_j \hat{a}_l \rangle,
\end{equation}
where $\langle \cdot\rangle$ represents the ensemble average.
Therefore, the ensemble average of the non-local atom correlation $K_{\boldsymbol \Delta}$ can be easily obtained by Fourier transformation.
For real experiments, two factors should be taken into account; an interaction effect and the finite-TOF effect.
A careful estimation of our experimental conditions~\cite{nakamura-19} shows that the ratio of the interaction energy $Un(n-1)/2$ to the kinetic energy  $\hbar \omega_L$ is mostly far lower than 1, justifying our ignorance of the interaction effect during TOF.
The finite-TOF effect is small but not negligible so that we determined the non-local atom correlation by extrapolation based on the theoretical model described in ref.~\cite{nakamura-19}.

\subsection*{Measurement of ensemble average of interaction energy}
In order to measure the ensemble average of the interaction energy $(1/2)U\sum_i \langle \hat{a}^{\dagger}_i \hat{a}^{\dagger}_i \hat{a}_i \hat{a}_i\rangle$ = $(1/2)U\sum_i \langle \hat{n}_i(\hat{n}_i-1)\rangle$, a method for projecting the distribution of the atom-number per site on an observable, namely, the atom-number-projection method, is required.
Details are described in ref.~\cite{nakamura-19}.
First, we increase the optical lattice depth quickly in order to freeze the hopping of atoms.
The ramp-up time is smaller than the hopping time, but large enough to prevent the atoms from being excited into the higher band of the optical lattice.
For example, the ramp-up time is $0.1$ ms from 5$E_{\rm R}$ to 15$E_{\rm R}$.

Subsequently, we perform a site-occupancy-resolved spectroscopy.
We employ two methods; the high-resolution spectroscopy using the optical transition between the $^{1}\text{S}_0$ and $^{3}\text{P}_2$ ($m_J=0$) electronic states of Yb atoms and the photoassociation spectroscopy.
The excellent resolution of the spectroscopy using the $^{1}\text{S}_0$ - $^{3}\text{P}_2$ ($m_J=0$) transition allows us to distinguish different site-occupancies, owing to quite different two-body interactions of $U_{eg}/h$ = -8.5 kHz and $U_{gg}/h$ = 3.2 kHz at 15$E_{\rm R}$.
From the area of the spectra, we obtain the total number $N_n$ of $n$-occupied sites.
The interaction energy is obtained as $U=(1/2)\sum_{n} N_n n(n-1)$.
The correlation factors induced by occupancy-dependent Rabi frequencies and a loss of atoms in the $^{3}\text{P}_2$ ($m_J=0$) state during the spectroscopy were studied in our previous work~\cite{nakamura-19}.

Another method that we use is the photoassociation which is a process to create one molecule from two atoms by light.
The created molecule rapidly escapes from the trap so that we can measure the total number of doubly occupied sites as the loss of the atoms. 
It is noted that the method is invalid in the case of triple and higher occupancies.
For example, photoassociation in a triply occupied site induces only two-atom loss and one atom remains, which is the same result as the case of a doubly occupied site, concerning the loss of atoms.

In our experiment, we employ the high-resolution spectroscopy using the optical transition between the $^{1}\text{S}_0$ and $^{3}\text{P}_2$ ($m_J=0$) for the 3D quench experiment.
In contrast, we use the photoassociation for the 1D quench experiment, where we additionally check the absence of triply occupied sites or higher by means of the high-resolution spectroscopy.



%

\section*{Acknowledgments}
We thank J. Sakamoto for experimental assistance and T. Kuno, Y. Watanabe and T. Sagawa for their comments and discussions.
The MPS-based calculations in Figs. 1, 2, and S6 were performed using the ITensor Library (http://itensor.org).
The TWA simulation in Fig. 4 was carried out at the Yukawa Institute Computer Facility.

This work was supported by the Grant-in-Aid for Scientific Research of the Ministry of Education, Culture Sports, Science, and Technology / Japan Society for the Promotion of Science (MEXT/JSPS KAKENHI) Nos. JP25220711, JP16H00801, JP17H06138, JP18H05405, JP18K03492, and JP18H05228; the Impulsing Paradigm Change through Disruptive Technologies (ImPACT) program; Japan Science and Technology Agency CREST (No. JPMJCR1673), and MEXT Quantum Leap Flagship Program (Q-LEAP) Grant Number JPMXS0118069021.

\clearpage



\renewcommand{\theequation}{S\arabic{equation}}
\renewcommand{\thesubsection}{\Roman{subsection}.}
\section*{Supplementary materials}

\subsection{Experimental setup and procedure}


Experimental procedures to load atoms into the optical lattice are shown in Fig.~\ref{fig:procedure}.
The beam waist of the horizontal FORT is about 15 $\mu$m and 33 $\mu$m.
The beam waist of the vertical FORT is about 43 $\mu$m and 126 $\mu$m.
The beam waist of the optical lattice is about 100 $\mu$m.

The optical lattice depth is calibrated by a pulsed optical lattice method (see also~\cite{kato-16}).
The accuracy of the calibration is typically within 5\%. 
In order to obtain the lattice parameters such as the on-site interaction energy $U$ and the tunneling-matrix element $J$, we use the numerical values of $U$ and $J$ calculated as functions of lattice depth reported in ref~\cite{krutitsky-16}.
The calculated lattice parameters are shown in Table.~\ref{tbl:UoJ}.

\subsection{Typical spectra and TOF images}

\subsubsection*{Three-dimensional case}

Figure~\ref{fig:spectra3D} shows the typical high-resolution spectra and the TOF images after the quench from $s=15$ to $s=5$ in the 3D lattice.
Figure~\ref{fig:spectra3D}A shows the high-resolution spectra and the TOF images just before the quench, that is, the adiabatically prepared state of
$s=15$.
The single peak of our spectra indicates that almost all sites are prepared at the $n=1$ Fock state.
The TOF image is also shown in the inset.
After the quench from $s=15$ to $s=5$ in 0.1 ms, the component with $n=2$ appears (Fig.~\ref{fig:spectra3D}B).
The difference of the TOF images in Figs.~\ref{fig:spectra3D}A and \ref{fig:spectra3D}B mainly comes from the change of the Wannier function: the shallower the lattice depth, the narrower the width of the Wannier function in momentum space.
After hold time of 0.5ms, the component with $n=3$ also appears (Fig.~\ref{fig:spectra3D}C), although the TOF image is similar to the one just after the quench.
These states are different from the adiabatically prepared state of $s=5$ state (Fig.~\ref{fig:spectra3D}D).
The TOF image shows sharp interference pattern, which reflects the existence of coherence over sites.
It is noted that $n=3$ sites appear in the case of the 3D quench from $s=15$ to $s=5$, which results in the fact that we need high-resolution spectroscopy to measure the interaction energy.
Interaction energy measurement with photoassociation is invalid for the $n \geq 3$ case because the photoassociation loss induces only the two-body loss.

\subsubsection*{One-dimensional case}

Figure~\ref{fig:spectra1D} shows the typical high-resolution spectra and the TOF images after the quench from $s=15$ to $s=5$ in the 1D lattice.
Figure~\ref{fig:spectra1D}A shows the high-resolution spectra and TOF images just before the quench and is the same as Fig.~\ref{fig:spectra3D}A.
Spectra and TOF images after the quench are shown in Figs.~\ref{fig:spectra1D}B and \ref{fig:spectra1D}C.
It is noted that multiple occupancy is suppressed compared to the 3D case and $n>2$ occupancy is negligible.
Therefore, the interaction energy measurement with photoassociation is valid in the case of the 1D case.

\subsection{Long-time behavior after the quench}

Long-time behaviors of the kinetic and interaction energies are shown in Fig~\ref{fig:atomnumber1D}A (1D case) and Fig~\ref{fig:atomnumber3D}A (3D case).
Note that the data for short-time periods are the same as those already shown in Fig. 1 for the 1D and Fig. 3 for the 3D in the main text, respectively.

Figures~\ref{fig:atomnumber1D}B and \ref{fig:atomnumber3D}B show the remaining atom numbers  after the quench from $s=15$ to $s=5$ in the 1D and 3D lattices, respectively.
The atom numbers after the quench are constant within the error-bars and the loss of atoms during the hold time is negligible.

Figures~\ref{fig:atomnumber1D}C and \ref{fig:atomnumber3D}C show the entropies after the quench from $s=15$ to $s=5$ in the 1D and 3D lattices, respectively.
The entropy is also almost constant and this means that heating and cooling during the hold time are negligible.
In order to measure the entropies, we adiabatically turned off the optical lattice and then measure the temperature in the optical trap.
The entropy $S$ in a harmonic trap is
\begin{equation}
S=4Nk_B\frac{\zeta(4)}{\zeta(3)}\left(\frac{T}{T_c}\right)^{3},
\end{equation}
where $N$ is the atom number, $T$ is the temperature, $T_c$ is a critical temperature for the Bose-Einstein condensation, $\zeta(z)$ is the zeta function, and $k_B$ is the Boltzmann constant.
\begin{equation}
k_B T_c= \hbar \bar{\omega} \left(\frac{N}{\zeta(3)}\right)^{1/3},
\end{equation}
where $\bar{\omega}$ is the geometric mean of the three trap frequencies, and $\hbar$ is the Planck constant divided by $2\pi$.
It is noted that the entropy before the quench is about $\sim 0.6 k_B$ and is much lower than the entropy after the quench because the non-adiabatic change of the lattice depth results in heating of the system.

\subsection{Numerical simulations in one dimension}

The numerical simulation data of the 1D system shown in the main text are obtained by the weighted averages of quantities from 1D tubes containing the different number of atoms.
The simulation of each 1D tube is performed by means of the time-dependent variational principle (TDVP) based on the matrix-product state representation of many-body wave functions~\cite{schollwock-2011,paeckel-19,haegeman-16}. 
The weights are determined on the basis of the local-density approximation (LDA).

In order to determine weights, we consider the Bose-Hubbard model (1) in a cubic lattice.
Since the initial lattice depth is sufficiently large and the number of particles at each site is not more than unity, we use the hard-core limit expression,
\begin{align}
\langle \hat{n}_i\rangle = 
\begin{cases} 
0, & \mu + V_i < -6J \\
\frac{1}{2} + \frac{\mu+V_i}{12J}, & -6J \leq \mu+V_i \leq 6J \\
1, & \mu + V_i > 6J
\end{cases}.
\end{align}
In a sufficiently large cubic lattice (we use 101 \(\times \) 101 \(\times \) 101 sites), we adjust \(\mu \) so that \(\sum_i \langle \hat{n}_i \rangle = 1.3 \times 10^4\) and count the number of 1D tubes with \(N\) atoms which we denote \(i_N\).
Then, we set a weight for quantities per particle from a tube with \(N\) atoms \(w_N\) to 
\begin{align}
w_N = \frac{i_N N}{1.3 \times 10^4}.
\end{align}
With \(V_i\) determined from the experimental setup, the largest \(N\) with finite \(i_N\) is 32.

For each 1D tube with \(N\) atoms, we simulate the quench dynamics by the two-site TDVP method~\cite{haegeman-16} following the procedure of the experiment: 
Taking the ground state of the Bose-Hubbard Hamiltonian (1) with \((V_x/E_{\rm R}, V_y/E_{\rm R}, V_z/E_{\rm R}) = (15, \infty, \infty)\) as an initial state, 
we perform the time evolution of the state with decreasing \((V_x/E_{\rm R}, V_y/E_{\rm R}, V_z/E_{\rm R})\) down to \((5, \infty, \infty)\) using the same time as that of the experiment.
We set the truncation error to be \(10^{-10}\) and the maximum occupation number of boson per site to be six.
The 1D tube used for numerical simulations consists of 48 sites.
We have confirmed that these parameters give sufficiently accurate results so that stricter parameters do not introduce significant changes within the presented time scale.

\subsection{Quench to a deeper Mott region in 1D}

In Figs. 2 and 4 of the main text, we show the correlation spreading after the quenches for the final lattice depth $s=5$ in 1D and $s=9$ in 2D, in which the phase velocity is not clearly separated from the group velocity. In Fig. S6, we show an example, in which the two velocities are well separated. Specifically, we depict the time evolution of the single-particle correlation function $K_{\Delta}(t)$ at $\Delta = 1, 2, 3,$ and $4$, where the final lattice depth is $s=9.4$ ($U/J = 25.3$) in 1D. The solid lines represent the numerical results by the MPS method, where we clearly see the multiple peaks. If we extract the propagation velocity from the first peak at each distance, which corresponds to the phase velocity, $v \simeq 20 Jd/\hbar$ and it is significantly larger than $v_{\rm max} = 6.0 Jd/\hbar$. By contrast, if we extract the propagation velocity from the highest peak of $|K_{\Delta}(t) – K_{\Delta}(0)|$ at each distance, which approximately corresponds to the group velocity, $v \simeq 5.7 Jd/\hbar$ and it is smaller than $v_{\rm max}$. This result also means that the behavior that the phase velocity can be larger than $v_{\rm max}$ is not a unique feature in 2D but it emerges due to the separation of the phase and group velocities at large $U/J$ regardless of the spatial dimension. Notice that similar physics has been already discussed in ref.~\cite{despres-19}.

\subsection{2D quench: peak and trough determination}

For determining the first peak and trough in the time evolution of the single-particle correlation function in the case of the 2D quench of the data shown in Fig. 4 of the main text, we assume the function as
\begin{equation}
f(t)=a_0+a_1\exp[\frac{-(t-t_0)^2}{2s^2}]+a_2\left[1-\exp(-\frac{t}{\tau})\right].
\end{equation}
The fitting results are shown in Fig.~\ref{fig:peakfit2d}.
The time of the first peaks (troughs) are numerically obtained from the fitting function.



\clearpage


\clearpage

\clearpage


\renewcommand{\thefigure}{S\arabic{figure}}
\renewcommand{\thetable}{S\arabic{table}}
\setcounter{figure}{0}
\setcounter{table}{0}

\begin{figure}[bt]
	\centering
	\includegraphics[width=7cm]{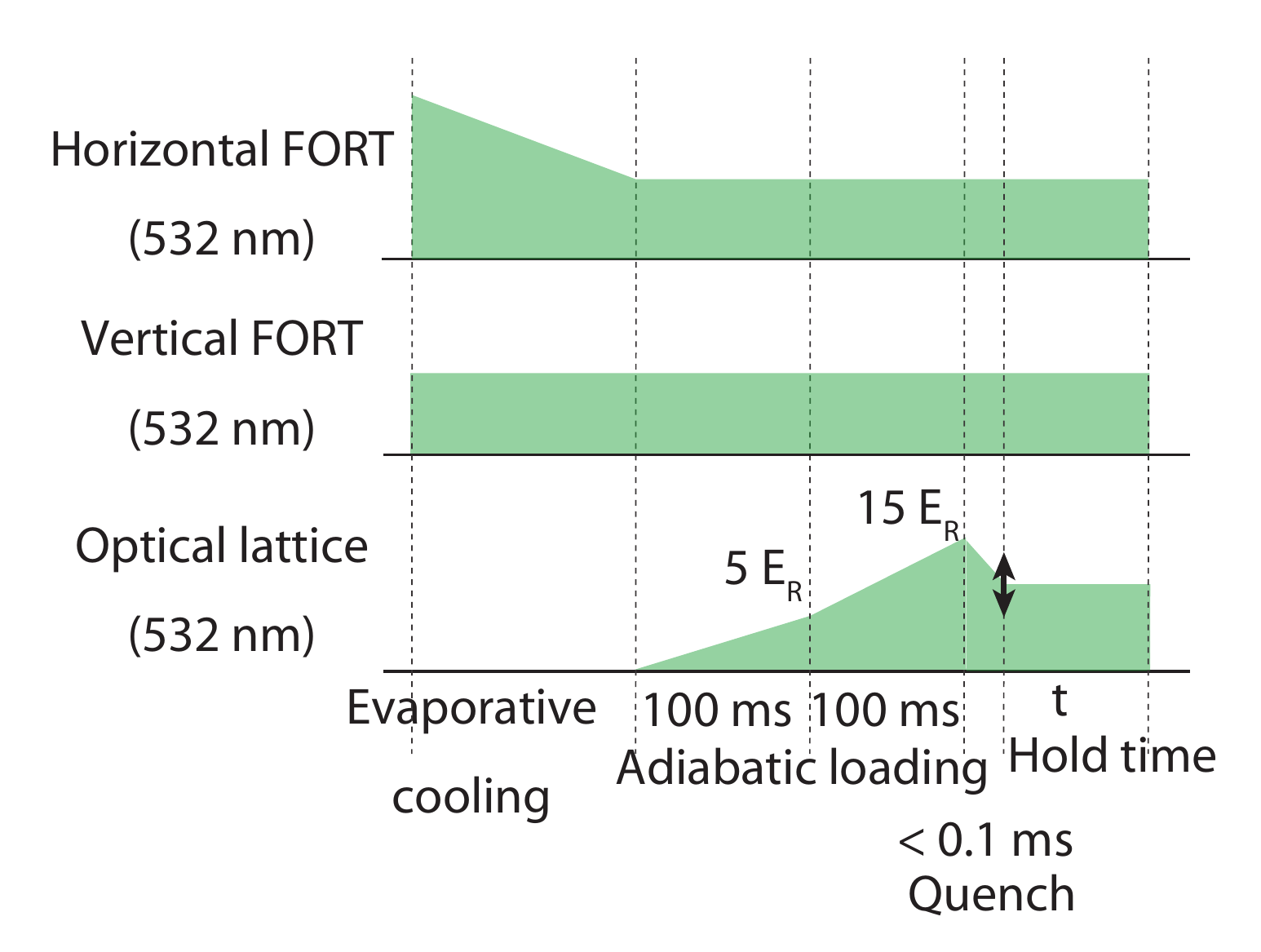}
	\caption{{\bf Time sequence for the preparation of an initial Mott state and the lattice quench.} The arrow shows a variable parameter of the lattice depth. \label{fig:procedure}}
\end{figure}

\begin{figure}[bt]
	\centering
	\includegraphics[width=7cm]{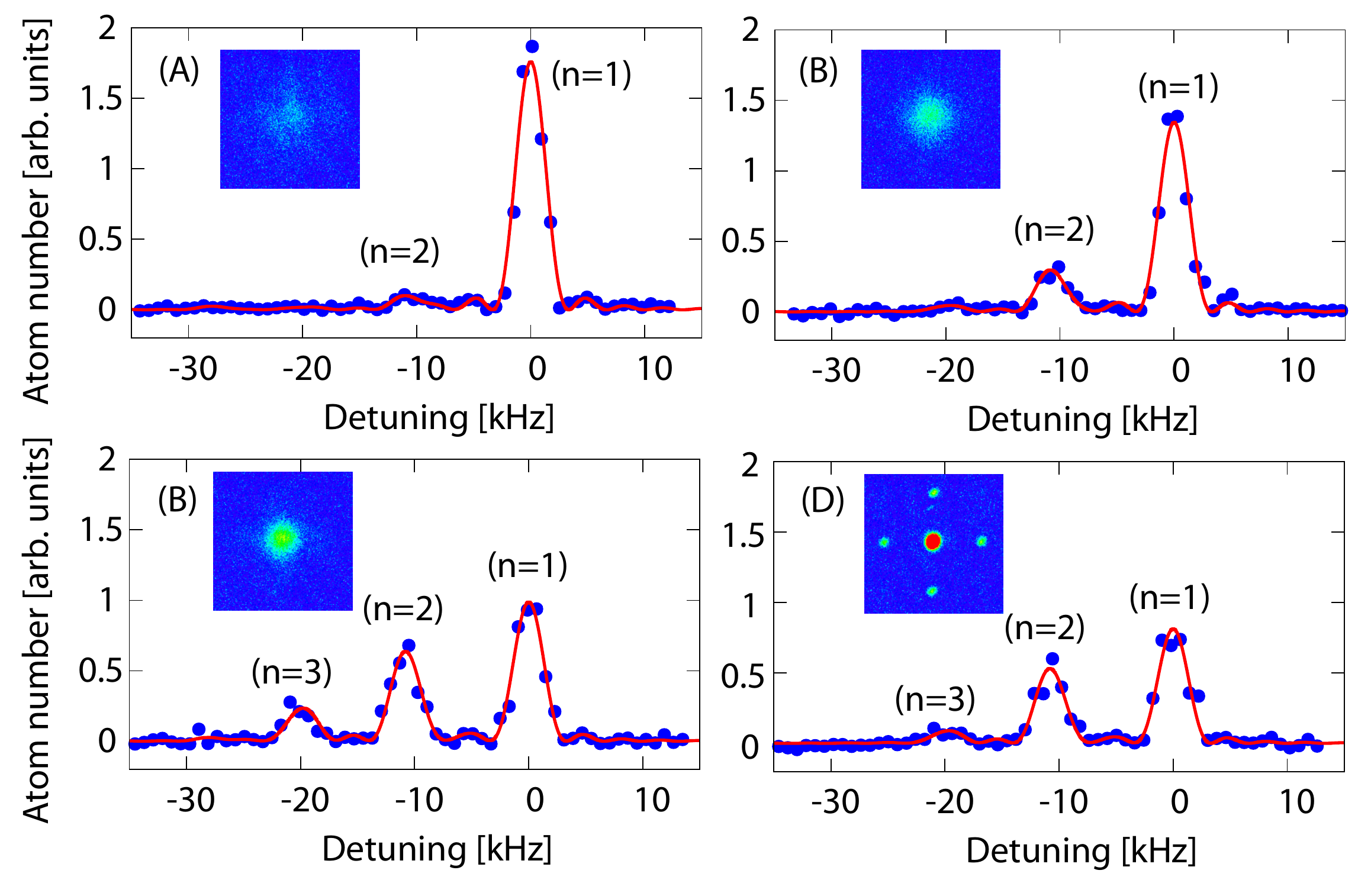}
	\caption{{\bf Typical high-resolution spectra and TOF images in the case of 3D.} ({\bf A}) before the quench ({\bf B}), ({\bf C}) after the quench. Hold time is ({\bf B}) 0 ms, and ({\bf C}) 0.5 ms. ({\bf D}) Adiabatic preparation. \label{fig:spectra3D}}
\end{figure}


\begin{figure}[bt]
	\centering
	\includegraphics[width=7cm]{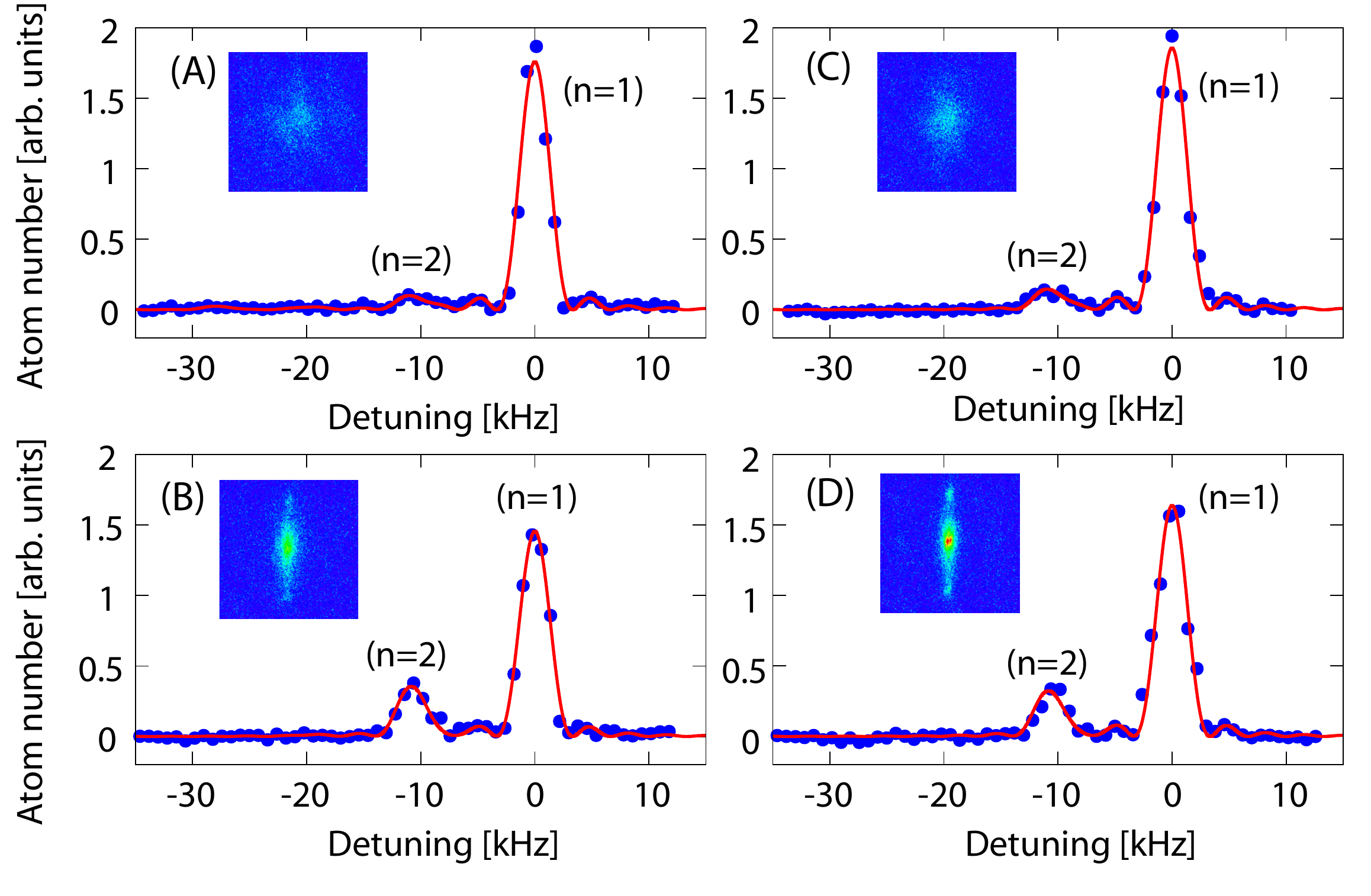}
	\caption{{\bf Typical high-resolution spectra and TOF images in the case of 1D.} ({\bf A}) before the quench ({\bf B}), ({\bf C}) after the quench. Hold time is ({\bf B}) 0 ms, and ({\bf C}) 0.5 ms. ({\bf D}) Adiabatic preparation. \label{fig:spectra1D}}
\end{figure}


\begin{figure*}[bt]
	\centering
	\includegraphics[width=14cm]{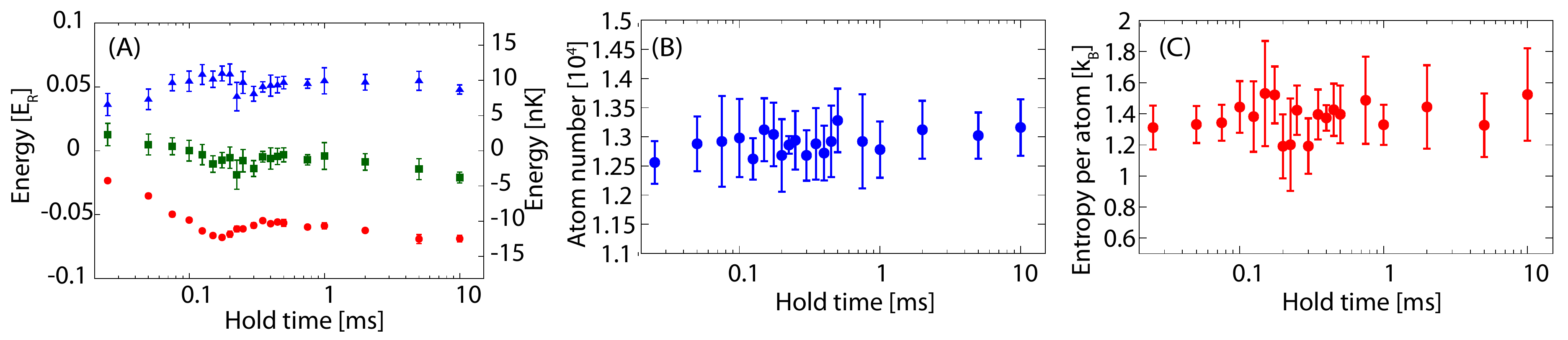}
	\caption{{\bf Long-time behaviors of the energy, the atom number, and the entropy after the 1D quench from $s=15$ to $s=5$ ($U/J=6.8$).} ({\bf A}) The kinetic-energy term (red), the onsite-interaction-energy term (blue), and the sum of them (green) are shown as functions of the hold time. ({\bf B}) The remaining atom number and ({\bf C}) the entropy after the quench as functions of the hold time.\label{fig:atomnumber1D}}
\end{figure*}


\begin{figure*}[bt]
	\centering
	\includegraphics[width=14cm]{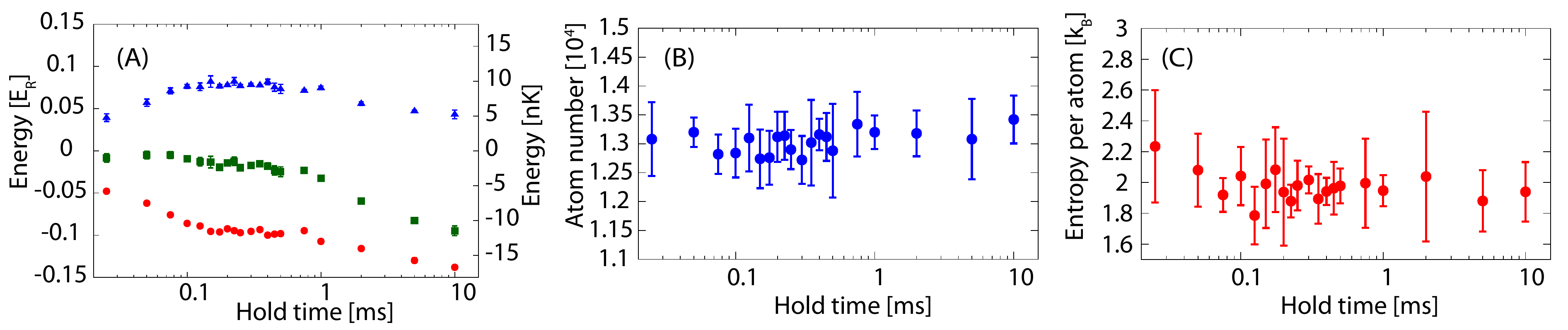}
	\caption{{\bf Long-time behaviors of the energy, the atom number, and the entropy after the 3D quench from $s=15$ to $s=5$ ($U/J=3.4$).} ({\bf A}) The kinetic-energy term (red), the onsite-interaction-energy term (blue), and the sum of them (green) are shown as functions of the hold time.  ({\bf B}) The remaining atom number and ({\bf C}) the entropy after the quench as functions of the hold time.\label{fig:atomnumber3D}}
\end{figure*}


\begin{figure*}[tb]
	\centering
	\includegraphics[width=14cm]{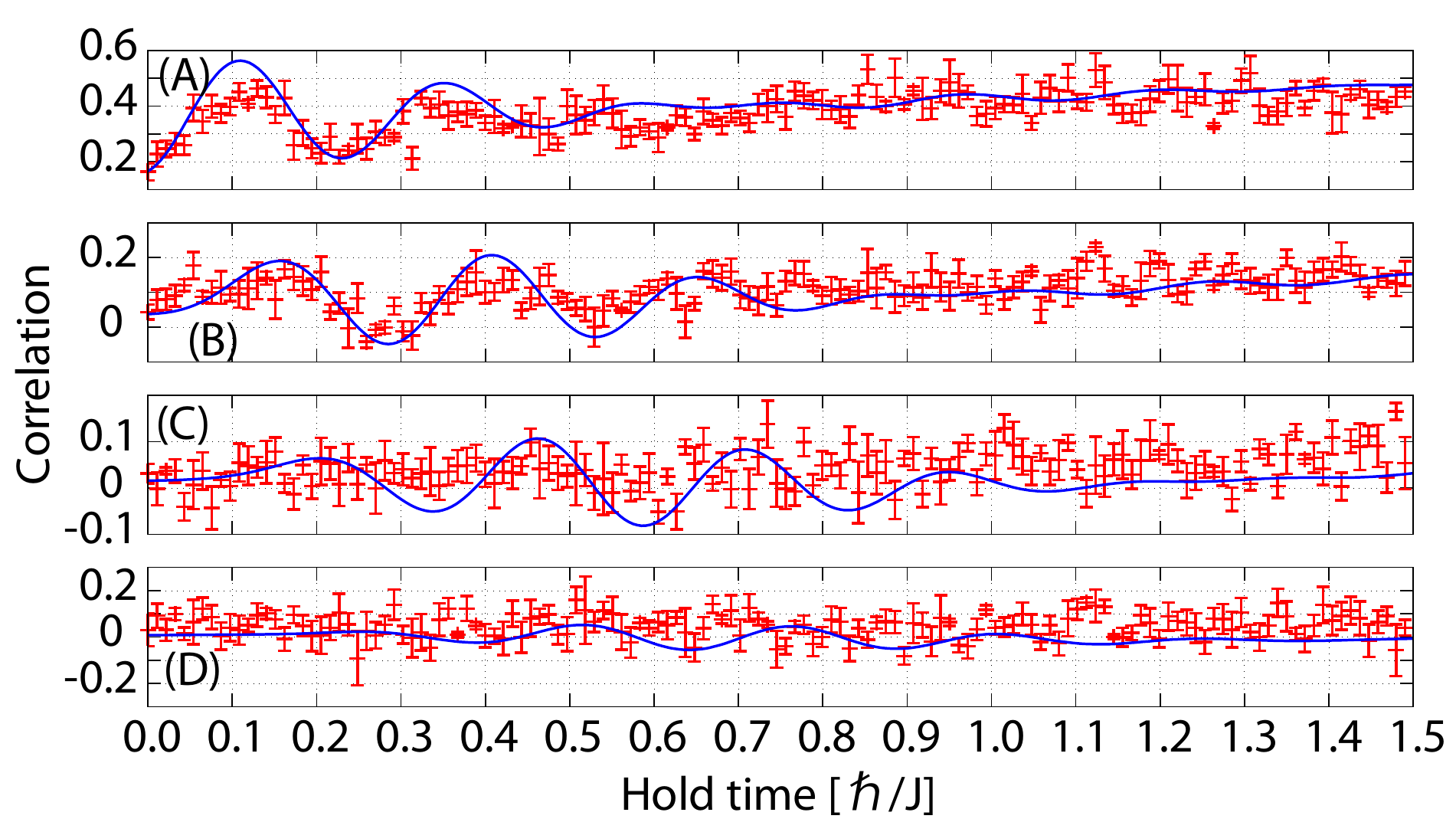}
	\caption{{\bf Experimental result and numerical study on the 1D quench from a deep Mott state to another deep Mott region.}
		The red points show experimental results with $U/J=25.3$ and the blue solid lines show the numerical calculations with $U/J=25$ obtained using the MPS method.
		({\bf A}) $\Delta=1$.
		({\bf B}) $\Delta=2$.
		({\bf C}) $\Delta=3$.
		({\bf D}) $\Delta=4$.	
		\label{fig:Correlation1Ddeep}}
\end{figure*}


\begin{figure*}[tb]
	\centering
	\includegraphics[width=14cm]{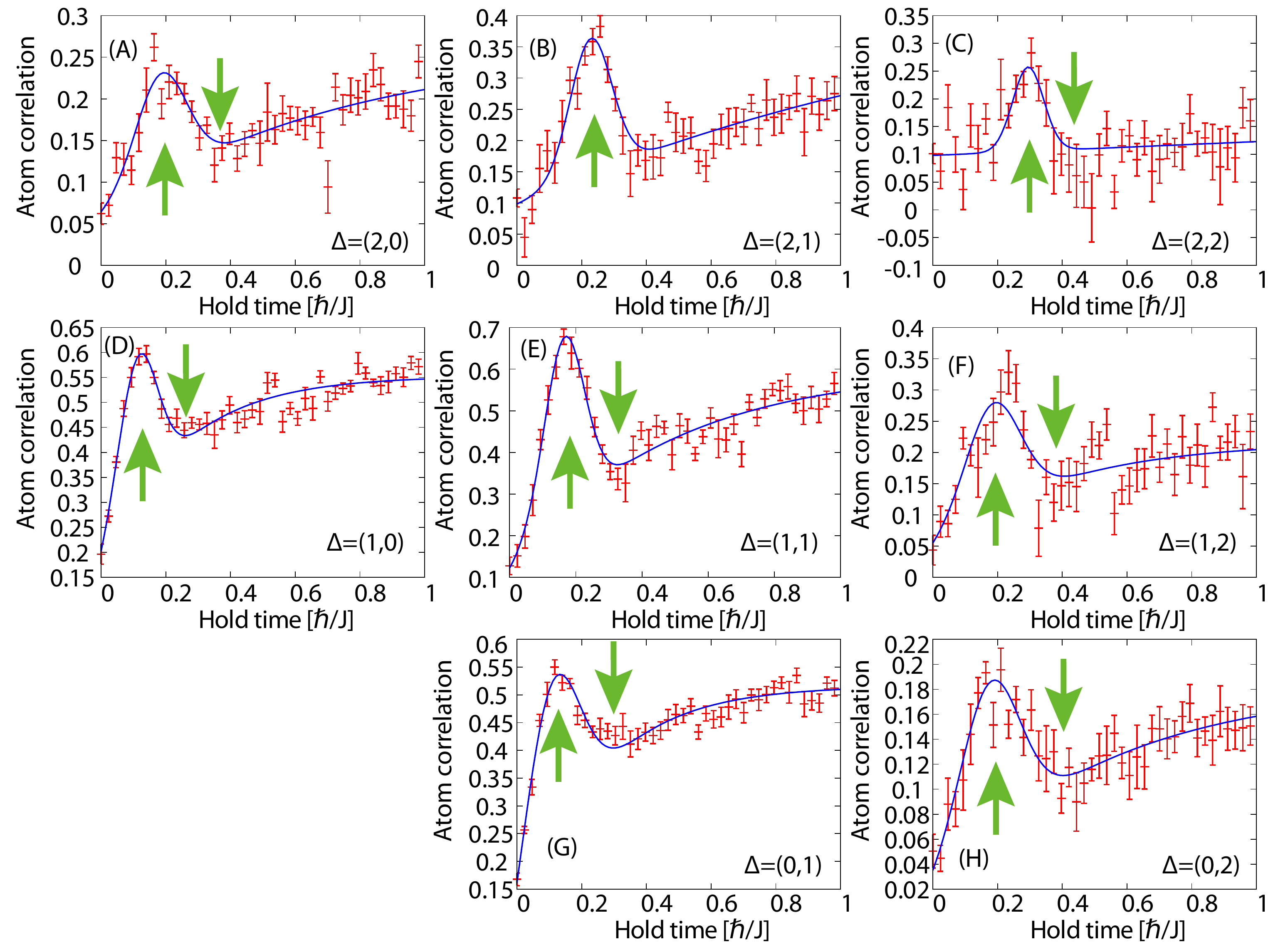}
	\caption{{\bf Peak and trough determination in the case of the 2D quench.} The fitting lines are shown by the blue solid lines.
		The upward (downward) arrows show the first peak (trough). 
		({\bf A}) ($\Delta x$, $\Delta y$)=(2,0),
		({\bf B}) ($\Delta x$, $\Delta y$)=(2,1),	
		({\bf C}) ($\Delta x$, $\Delta y$)=(2,2),
		({\bf D}) ($\Delta x$, $\Delta y$)=(1,0),
		({\bf E}) ($\Delta x$, $\Delta y$)=(1,1),
		({\bf F}) ($\Delta x$, $\Delta y$)=(1,2),
		({\bf G}) ($\Delta x$, $\Delta y$)=(0,1),
		({\bf H}) ($\Delta x$, $\Delta y$)=(0,2).
		\label{fig:peakfit2d}}
\end{figure*}


\begin{table}
\centering
		\begin{tabular}{c|ccc}
			Lattice depths [$E_{\rm R}$] & $U$ [$E_{\rm R}$] & $J$ [$E_{\rm R}$] &$U/J$ \\
			\hline
			(5, 15, 15)  & 0.45(1) & 0.066(4) & 6.8(5) \\ 
			(5, 5, 5) & 0.22(1) & 0.066(4) & 3.4(3) \\
			(9, 9, 15) & 0.47(1) & 0.024(3) & 19(2) \\ 
			(9.4, 15, 15) & 0.55(1) & 0.0220(2) &25(3) \\
			(15, 15, 15) & 0.64(2) & 0.0065(9) & 98(14) 
		\end{tabular}
	\caption{{\bf On-site interaction energies and tunneling-matrix elements.} The lattice depths, $U$, and $J$ are shown in units of $E_{\rm R}$. The largest tunneling-matrix elements are only shown. The uncertainties of the parameters are calculated on the assumption that the accuracy of our determination on the lattice depth is 5\%. \label{tbl:UoJ}}
\end{table}

\end{document}